\documentclass[structabstract]{aa} 
\usepackage{graphicx} 
\usepackage{txfonts} 
\usepackage{natbib} 
\usepackage{epstopdf}
\bibpunct{(}{)}{;}{a}{}{,}
\begin{document}
\title{Stellar mass map and dark matter distribution in M\,31}
\titlerunning{Visible and dark mass of M\,31}
\author{A. Tamm\inst{1} 
\and E. Tempel\inst{1,}\inst{2} 
\and P. Tenjes\inst{1,}\inst{3}
\and O. Tihhonova\inst{1,}\inst{3}
\and T. Tuvikene\inst{1}}
\institute{Tartu Observatory, Observatooriumi~1, 61602 T\~oravere, Estonia\\
    \email{[atamm;elmo]@aai.ee}
\and National Institute of Chemical Physics and Biophysics, R\"avala pst 10, Tallinn 10143, Estonia
\and Institute of Physics, University of Tartu, Riia 142, Tartu 51014, Estonia}
\date{Received 00 Month 2012 / Accepted 00 Month 2012}

\abstract
%context
{}
%aims
{Stellar mass distribution in the Andromeda galaxy (M\,31) is estimated using optical and near-infrared imaging data. Combining the derived stellar mass model with various kinematical data, properties of the dark matter (DM) halo of the galaxy are constrained.}
%methods
{SDSS observations through the $ugriz$ filters and the Spitzer imaging at 3.6 microns are used to sample the spectral energy distribution (SED) of the galaxy at each imaging pixel. Intrinsic dust extinction effects are taken into account by using far-infrared observations. Synthetic SEDs created with different stellar population synthesis models are fitted to the observed SEDs, providing estimates for the stellar mass surface density at each pixel. The stellar mass distribution of the galaxy is described with a 3-dimensional model consisting of a nucleus, a bulge, a disc, a young disc and a halo component, each following the Einasto density distribution (relations between different functional forms of the Einasto density distribution are given in Appendix~\ref{app:2}). By comparing the stellar mass distribution to the observed rotation curve and kinematics of outer globular clusters and satellite galaxies, the DM halo parameters are estimated.}
%results
{Stellar population synthesis models suggest that M\,31 is dominated by old ($\ga7$\,Gyr) stars throughout the galaxy, with the lower limit for the stellar mass-to-light ratios $M/L_r\ga 4\,M_{\odot}/L_{\odot}$. The upper limit $M/L_r\la 6\,M_{\odot}/L_{\odot}$ is given by the rotation curve of the galaxy. The total stellar mass is (10--$15)\cdot10^{10}M_{\odot}$, 30\% of which is in the bulge and 56\% in the disc. None of the tested DM distribution models (Einasto, NFW, Moore, Burkert) can be falsified on the basis of the stellar matter distribution and the rotation curve of the galaxy. The virial mass $M_{200}$ of the DM halo is $(0.8$--$1.1)\cdot10^{12}M_{\odot}$ and the virial radius is $R_{200}=$189--213~kpc, depending on the DM distribution. For the Einasto profile, the average density of the DM halo within the central 10 pc is 16--61\,$M_{\odot}\mathrm{\,pc^{-3}}$ (0.6--2.3\,$\mathrm{TeV}/c^2\,\mathrm{cm}^{-3})$, depending on the stellar mass model. The central density of the DM halo is comparable to that of nearby dwarf galaxies, low-surface-brightness galaxies and distant massive disc galaxies, thus the evolution of central DM halo properties seems to be regulated by similar processes for a broad range of halo masses, environments, and cosmological epochs.}
%conclusions - vabatahtlik
{}


\maketitle

%=========================================================
\section{Introduction}
%=========================================================

Due to its proximity and size, our nearest large neighbour galaxy M\,31 offers a unique and attractive opportunity to study stellar populations and galactic structure in detail. It has been a source for groundbreaking discoveries ever since the secure acknowledgement of this nebula as an extragalactic object by Ernst \"{O}pik \citeyearpar{opik:22}.

Self-consistent treatment of the available photometry and kinematical data of M\,31 enabled the construction of sophisticated multi-component galactic models already decades ago \citep[e.g.][]{tenjes:94}. Considering the huge volume and high detail of observational information available nowadays, complex mass models of M\,31 offer a promising opportunity for casting light on one of the most puzzling problems in astrophysics and cosmology: the nature and properties of dark matter (DM) haloes. By now, the increasing scope of observations has enabled stretching mass models much further than the extent of gas disc rotation curves, providing new clues about DM halo parameters \citep{geehan:06,seigar:08,chemin:09,corbelli:10}.

On the other hand, particle physics instrumentation has reached a level at which it can provide some hints about DM. Although the diffuse Galactic background likely exceeds the expected flux of high-energy particles resulting from decaying or annihilating DM in extragalactic sources \citep{bertone:07,Hutsi:10}, particles from more concentrated extragalactic objects might be detectable as an enhancement of the Galactic signal within certain apertures. By comparing the assumed DM distribution in M\,31 to the data collected with the diverse arsenal of ground-based and space-borne detectors of high-energy particles, some constraints on the energy spectrum of DM particles have already been laid \citep[e.g.][]{aharonian:03,lavalle:06,Boyarsky:08,dugger:10,Watson:12}.

For more stringent constraints, not only more capable detectors are needed but also a better understanding of the properties of the source DM haloes. As bizarre as it seems, the derivation of the detailed mass distribution of the \object{Andromeda} galaxy was limited by the lack of suitable optical imaging up to recent times.

Although visible even to the naked eye, its span over four degrees on the celestial sphere makes Andromeda a real challenge to observe with a usual scientific telescope. Thus it is only very recently that observations covering the galaxy with deep wide-field CCD imaging have started to appear: a dedicated scan within the Sloan Digital Sky Survey \citep{york:00}, the Canada-France-Hawaii telescope Megacam programme PAndAS \citep{mcconnachie:09}, the Pan-STARRS telescope project PAndromeda \citep[][]{lee:12}. Combined with the space-based ultraviolet \citep{thilker:05}, near- \citep{barmby:06} and far-infrared \citep{gordon:06,fritz:11} observations, these data now provide an unprecedented panchromatic view of a galaxy at a resolution of a few parsecs, allowing for the derivation of detailed properties of the stellar populations.

In this paper we estimate the stellar and DM distribution of M\,31 using the SDSS and Spitzer 3.6-micron imaging to constrain the properties of the stellar populations. We construct a mass distribution model of the galaxy in correspondence with the latest kinematical data from the literature, giving estimates for the DM halo properties and the related uncertainties.

We have taken the inclination angle of M\,31 to be 77.5\degr \citep{Walterbos:88,deVaucouleurs:91} and the distance to the galaxy \mbox{785\,kpc} \citep{McConnachie:05}, yielding the scale \mbox{228\,pc/arcmin}. Throughout the paper, luminosities are presented in AB-magnitudes and are corrected for the Galactic extinction according to \citet{tempel:11}, where extinction corresponding to the Sloan filters is derived from the \citet{Schlegel:98} estimates and the Galactic extinction law by linear interpolation. The absolute solar luminosity for each filter is taken from \citet{blanton:07}.

%=========================================================
\section{Observational SEDs}  \label{sect:obs}
%=========================================================

In an ideal case, one would study stellar populations using all the available photometric data to sample the spectral energy distribution (SED) of a galaxy throughout the full electromagnetic spectrum. However, we have limited ourselves here to the optical and near-infrared section of the spectrum, since its interpretation with synthetic stellar population models is most straightforward. Also the stellar mass is best traced by this wavelength domain.

For deriving the observed SEDs of M\,31, we relied on the Sloan Digital Sky Survey (SDSS) observations through the $ugriz$ filters and the Spitzer Space Telescope IRAC camera imaging at 3.6 microns. For our purposes, these observations provide a sufficiently wide and deep coverage and the calibration of the data is relatively well-understood.

The basic steps of the SDSS image processing and mosaicing used here have been introduced in \citet{tempel:11}. The intrinsic absorption of the galaxy has been taken into account by applying the dust disc model developed in \citet{tempel:10,tempel:11} on the basis of the far-infrared flux distribution as measured by the Spitzer MIPS camera. We have used the resultant absorption-free SDSS images for recovering the total starlight along sight-lines affected by the dust disc. The final SDSS mosaic was resampled to 3.96\,arcsec\,px$^{-1}$. 

To reconstruct the near-infrared view of M\,31 we retrieved the pipeline-calibrated and -processed (post-BCD) Spitzer images from the NASA/IPAC Infrared Science Archive. Exposures severely suffering from cosmic rays were omitted. A mosaic image was created using pointing information in the image headers. The final mosaic was resampled to the same pixel scale as the SDSS images.

The spatial resolution of the dust absorption model was limited by the point-spread-function (PSF) of the Spitzer MIPS camera at 160 microns, thus we convolved the SDSS and Spitzer images with the same PSF. This step improved also the signal-to-noise ratio in the outer regions of the galaxy and removed negative pixel values resulting from the noise after sky removal.

The images were matched with each other geometrically to sub-pixel accuracy. Foreground stars, satellite galaxies and background objects were masked with a common mask for each filter, constructed using SExtractor \citep{Bertin:96} and manual region placing. Additionally, the noisy edges of the Spitzer images were masked. Now the SED could be directly derived by calibrating the intensity within each pixel to standard flux units.

%=========================================================
\section{Model SEDs} \label{sect:mod}
%=========================================================

A wide variety of synthetic stellar population models is available for interpreting the observed spectral energy distributions. In order to address possible degeneracies and uncertainties of such models, we have used three models to reproduce the observational SEDs. The chosen models follow significantly different approaches for generating the properties of the synthetic stellar populations: (I) the composite model spectra by \citet[][hereafter B07]{blanton:07} are composed as a mixture of \citet{bruzual:03} instantaneous-burst stellar population models of different ages and metallicities, and models of gas emission from MAPPINGS-III \citep{kewley:01}; (II) \citet[][hereafter M05]{maraston:05} models lay a special emphasis on the thermally pulsing asymptotic giant branch stars; (III) the evolutionary synthesis model GALEV \citep{kotulla:09} is the sole model in which the chemical evolution of gas and stars is treated self-consistently. In B07, the \citet{chabrier:03} stellar initial mass function (IMF) was used; for M05 and GALEV, we have chosen the \citet{kroupa:01} IMF option.

B07 provides five composite spectra corresponding to an extremely young, an old, and three intermediate populations. It is shown in B07 that a linear combination of these spectra can adequately describe the spectral energy distribution of most of the galaxies. However, this aspect alone does not necessarily prove that the underlying models are meaningful; rather, it demonstrates that the spectra are sufficiently diverse. Thus for the M05 and GALEV models, we have followed the general observational knowledge about the star formation history and metallicity of M\,31 to tame the age-metallicity degeneracy, assuming that much of the galaxy is dominated by old stars of nearly solar metallicity, while the star-forming ring is composed of younger stars \citep{bellazzini:03,sarajedini:05,brown:06,olsen:06,saglia:10,zou:11}. From the available M05 models, we have selected single-, instantaneous-burst stellar populations. GALEV is used in the chemically consistent regime to generate old populations with different star formation histories, with and without an additional starburst having occurred 1--4 billion years ago to mimic the star-forming regions.

For each set of model spectra, we sought linear combinations of up to 5 spectra to represent the SED of M\,31 within each pixel according to the formula
\begin{equation}
f(\lambda)_{\mathrm{obs}} = \sum\limits_{i}{m_i f(\lambda)_i} ,
\label{eq:sed_fit}
\end{equation}
where $f(\lambda)_{\mathrm{obs}}$ is the observed SED, $f(\lambda)_i$ are the model SEDs per unit mass and $m_i$ are the corresponding weights of the model SEDs. In this formalism, $m_i$ effectively measure the mass of each model stellar population within a given pixel.

\begin{figure}
\includegraphics[width=88mm]{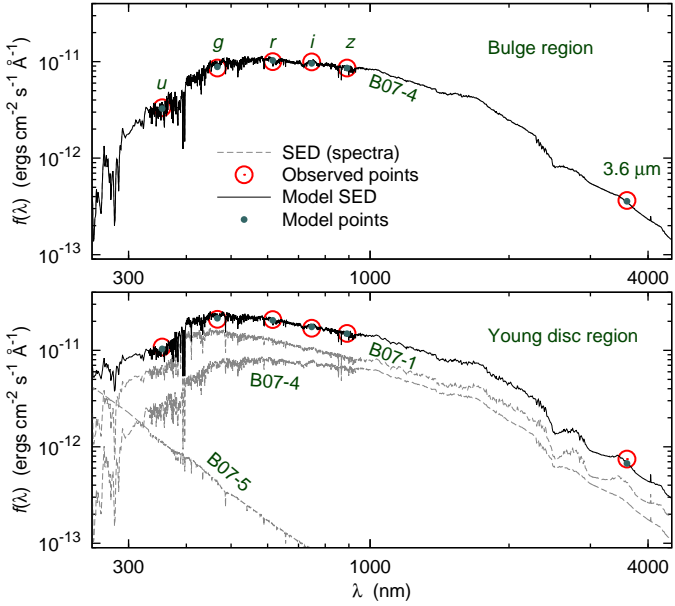}
\caption{Examples of the observed (large circles) and modelled (lines) SED for a random pixel in the bulge region (\emph{upper panel}) and in the young disc region (\emph{lower panel}). The sizes of the datapoints indicate the photometric uncertainties of each measurement. The model values corresponding to each filter are also shown (small datapoints). In most pixels, the reddest model population (B07-4) alone provides a good representation of the observed SED. In the young disc regions, the stellar populations are more diverse: in the lower panel, the B07 model populations 1, 4 and 5 contribute 24.44\%, 75.52\%, and 0.04\% of the mass, respectively. The corresponding SEDs are weighted according to the mass fraction. In this plot, the observed SEDs and the sum of the weighted model spectra are normalised per 1\,$M_{\odot}$ at a distance of 10\,pc.}
\label{fig:sed_fit}
\end{figure}

The spectra with a non-negligible contribution to the integral SED of the galaxy are listed in Table~\ref{tab:mod_spec} together with other relevant information. An illustration of the SED fitting within random pixels from the bulge and disc regions of the galaxy is presented in Fig.~\ref{fig:sed_fit}. Here, the observed flux through each of the six filters is overlaid with the best-fitting linear combination of the B07 composite spectra. 

\begin{table}
\caption{Synthetic stellar populations used for SED fitting.}
\begin{flushleft}
\begin{tabular}{lllllll}
\hline\hline
Name & Age &  [Fe/H] &  $\frac{M_\mathrm{tot}}{L_g}$ & $\frac{M_\mathrm{tot}}{L_r}$ & $\frac{M_\mathrm{tot}}{L_i}$ & Fract. \\
 &  [Gyr] &  & $[\frac{M_\odot}{L_\odot}]$ & $[\frac{M_\odot}{L_\odot}]$ & $[\frac{M_\odot}{L_\odot}]$ &  \\
(1) & (2) & (3) & (4) & (5) & (6) & (7) \\
\hline
B07-1  &  0.7     & 0.40  & 0.76   & 0.78   & 0.72   & 0.014 \\
B07-3  &  0.4--1  & 0.05  & 0.47   & 0.50   & 0.56   & 0.003  \\
B07-4  &  7--12   & 0.03  & 5.05   & 3.87   & 3.12   & 0.983  \\
\hline
M05-1  &  1     & 0.00  & 1.11   & 1.00   & 0.85   & 0.008 \\
M05-2  &  2     & 0.00  & 2.18   & 1.70   & 1.43   & 0.002 \\
M05-3  &  4     & 0.00  & 3.99   & 3.03   & 2.56   & 0.214  \\
M05-4  &  12    & 0.00  & 11.6   & 8.08   & 6.47   & 0.767  \\
M05-5  &  12    & -0.33  & 9.00   & 6.60   & 5.37   & 0.009  \\
\hline
GALEV-1  &  1, 10    & 0.04  & 2.88   & 3.14   & 2.92   & 0.004 \\
GALEV-2  &  2, 11    & 0.07  & 4.35   & 4.13   & 3.65   & 0.011 \\
GALEV-3  &  4, 13    & 0.09  & 7.58   & 6.20   & 5.23   & 0.089  \\
GALEV-4  &  12       & 0.12  & 4.63   & 4.55   & 4.05   & 0.015  \\
GALEV-5  &  12       & 0.18  & 10.9   & 8.33   & 6.86   & 0.881  \\
\hline
\end{tabular}
\end{flushleft}
\tablefoot{
The columns contain the following: (1) stellar population model; for B07 models the number is as in the original paper; (2) approximate age of the dominant star-formation epoch(s); (3) average metallicity of the stars; (4)--(6) mass-to-light ratio in the $gri$ filters; (7) total stellar mass fraction in M\,31 of the corresponding stellar population.}
\label{tab:mod_spec}
\end{table}

Following Eq.~(\ref{eq:sed_fit}), the SED-fitting process yielded the stellar mass of each model population within each imaging pixel. For every population synthesis model, the contribution of different synthetic populations to the integral mass of the galaxy is presented in the last column of Table~\ref{tab:mod_spec}. It is seen that for each set of model spectra, the reddest spectrum (corresponding to an old population with near-solar metallicity) dominates all across the galaxy and dictates its population properties. It is only within the star-forming ring that other spectra provide detectable contribution.

The mass-to-light ratios and thus the masses of the stellar components predicted by different population synthesis models are remarkably different, as indicated in Table~\ref{tab:mod_spec}. M05 and GALEV give much more massive stellar populations than B07. The scatter of the mass-to-light ratios of different models results from differences in the modelling approach and hence reflects the uncertainties of stellar mass estimations in general. Although the B07 model spectra provide the most precise description of the actual SEDs, we have no grounds to state that the B07 model represents the actual populations better than the M05 and GALEV models. Thus in the following, we have considered this mass scatter as an uncertainty of the final mass estimates. We used the B07 model for the lowest limit and planned initially to use the other models for the upper limit of the stellar mass. As shown below, however, the masses suggested by the M05 and GALEV models are contradicting the rotation curve measurements, which set a more rigid upper limit for the stellar masses. Nevertheless, we discourage the reader to use this result alone for judging the M05 and GALEV models in general. We can just conclude that with our currently used configuration of the stellar population details (i.e. the IMF, metallicity, and star formation history), these models overestimate the realistic stellar mass of M\,31.

%=========================================================
\section{Stellar mass distribution}\label{sect:stellar_m}
%=========================================================

\begin{figure}
\includegraphics[width=88mm]{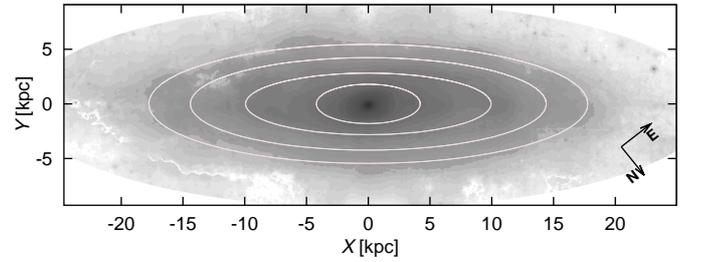}
\caption{Stellar mass-density map of M\,31. The ellipses enclose 50\%, 75\%, 90\%, and 95\% of the total mass, respectively.}
\label{fig:map}
\end{figure}

In the previous section, the stellar mass corresponding to each model spectrum was derived for each imaging pixel. This gives the 2-dimensional stellar mass distribution in M\,31, as presented in Fig.~\ref{fig:map} for the B07 model. The distribution appears featureless and regular, resulting from the intensive smoothing with the PSF of the Spitzer 160-micron imaging, but it indicates also that the galaxy is generally undisturbed and that the intrinsic dust-absorption effects have been removed in proper proportions along different lines of sight. 

The actual spatial distribution of stellar matter can be split into the contributions of different galactic components. We measured the elliptically averaged stellar mass distribution from Fig.~\ref{fig:map} and approximated it as a superposition of the stellar components: a nucleus, a bulge, a disc, and a halo. In addition, the ring-like star-forming region was taken as a separate component from the disc and is referred here as the young disc. Assuming each component to be an ellipsoid of rotational symmetry and constant axial ratio $q$, we used the Einasto law

\begin{equation}
\rho(a)= \rho_c\exp\left\{-d_{N}
\left[\left(\frac{a}{a_\mathrm{c}}\right)^{1/N}-1\right]\right\}
\label{eq:einasto}
\end{equation}
to describe the density distribution of a component. Here distance from the centre $a=\sqrt{r^2+z^2/q^2}$, where $r$ and $z$ are the two cylindrical coordinates; $d_N$ is a function of $N$, such that $\rho_c$ becomes the density at distance $a_{\mathrm{c}}$, which defines the volume containing half of the total mass of the component. The derivation of $d_N$ is presented in Appendix \ref{app:2}. Being mathematically identical to the S\'ersic's $R^{1/n}$ model but fitted to the space density, Eq.~(\ref{eq:einasto}) provides a sufficiently flexible distribution law for describing relaxed galactic components.

For a more accurate description of the star-forming region, the spatial density distribution of the young disc is assumed to have a toroidal form, approximated as a superposition of a positive and a negative density component, both following Eq.~(\ref{eq:einasto}) (see the last paragraph of Appendix~\ref{app:2} for more details).

The structural parameters of all the stellar components have been adopted from a previous analysis \citep{tempel:11}, also based on the SDSS data. In the referred paper, the Einasto law was expressed with respect to the harmonic mean radius instead of $a_c$. Relations between different functional forms of the Einasto distribution are presented in Appendix~\ref{app:2}.

%=========================================================
\begin{table*}
    \caption{Parameters of stellar components.}
    \centering
    \begin{tabular}{lcccccccccc}
        \hline\hline
        Component & $a_0$\tablefootmark{a}
                     & $a_c$ & $q$ & $N$ & $d_N$ & $\rho_c$ & $M_\mathrm{comp}$\tablefootmark{b} & $M/L_g$\tablefootmark{b} & $M/L_r$\tablefootmark{b} & $M/L_i$\tablefootmark{b} \\
             & [kpc] & [kpc] &     &     &  & $[M_\odot\,\mathrm{pc}^{-3}]$ & $[10^{10}M_\odot]$ & $[M_\odot/L_\odot]$ & $[M_\odot/L_\odot]$ & $[M_\odot/L_\odot]$  \\
        \hline
        Nucleus      & 0.01 & 0.0234 & 0.99 & 4.0 & 11.668 & $1.713\cdot 10^{0}$ & 0.008& 4.44 & 3.20 & 2.35 \\
        Bulge        & 0.63 & 1.155  & 0.72 & 2.7 &  7.769 & $9.201\cdot 10^{-1}$ & 3.1 & 5.34 & 4.08 & 3.01 \\
        Disc         & 7.7 & 10.67  & 0.17 & 1.2 &  3.273 & $1.307\cdot 10^{-2}$ & 5.6  & 5.23 & 3.92 & 2.92 \\
        Young disc\tablefootmark{c}
                     & 10.0 & 11.83  & 0.01 & 0.2 &  0.316 & $1.179\cdot 10^{-2}$ & 0.1 & 1.23 & 1.12 & 0.88 \\
        Stellar halo & 6.3  & 12.22  & 0.50 & 3.0 &  8.669 & $4.459\cdot 10^{-4}$ & 1.3 & 6.19 & 4.48 & 3.25 \\
        \hline
    \end{tabular}
    \tablefoot{
    Structural parameters $a_0$, $q$, $N$, and galaxy luminosities in SDSS filters ($L_g$, $L_r$, $L_i$) are taken from \citet{tempel:11}. Component masses $M_\mathrm{comp}$ are derived in this paper.
    \tablefoottext{a}{Harmonic mean radius (see Appendix~\ref{app:2}).}
    \tablefoottext{b}{Masses and mass-light-ratios corresponding to the B07 model, i.e. the lower limits; the upper limits (from the maximum-stellar model) are 1.5 times higher in each case.}
    \tablefoottext{c}{The structural parameters and $\rho_c$ are given for the positive component. In the dynamical models, the gas mass $6\cdot10^9M_{\odot}$ is added to the young disc.}
    }
 \label{table:stellar_comp}
\end{table*}
%=========================================================

\begin{figure}
\includegraphics[width=88mm]{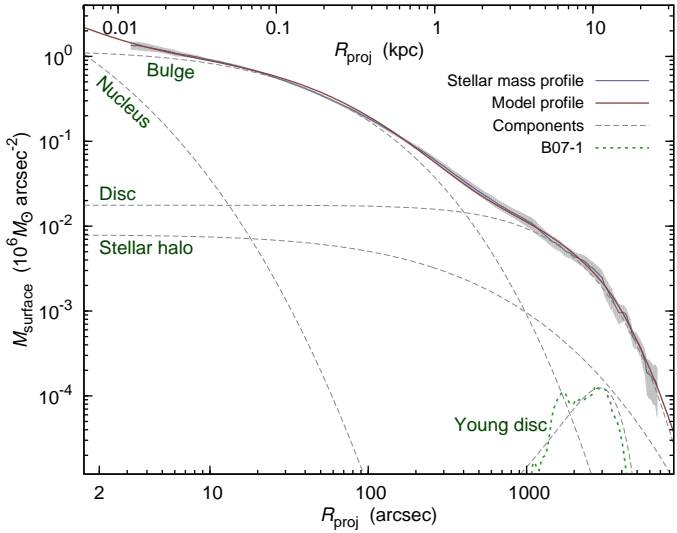}
\caption{The mass-density distribution of the galaxy, averaged along elliptical iso-density contours, as inferred from the B07 model (thick grey line; its thickness indicates deviations along each ellipse), the model profile (solid line) and the contributions of the individual stellar components (dashed lines) to the model profile. The contribution of the model population B07-1 to the mass distribution is also shown; it is closely traced by the young disc component of the model.}
\label{fig:comps}
\end{figure}

Using the previously derived structural parameters and leaving the component masses as free variables, we fitted these components to the elliptically averaged stellar mass distribution. The lower limit estimates of the masses of the stellar components, derived from the B07 model, are presented in Table~\ref{table:stellar_comp} together with other related parameters. The upper mass limits are constrained by the rotation curve and are 1.5 times higher for each component (see below). The corresponding mass distribution of each stellar component as well as the total stellar mass distribution of the galaxy are shown in Fig.~\ref{fig:comps} for the B07 model. To illustrate the correspondence between the young disc model component and the first model spectrum (B07-1), the contribution of the latter to the overall mass distribution is also shown.

\begin{figure}
\includegraphics[width=88mm]{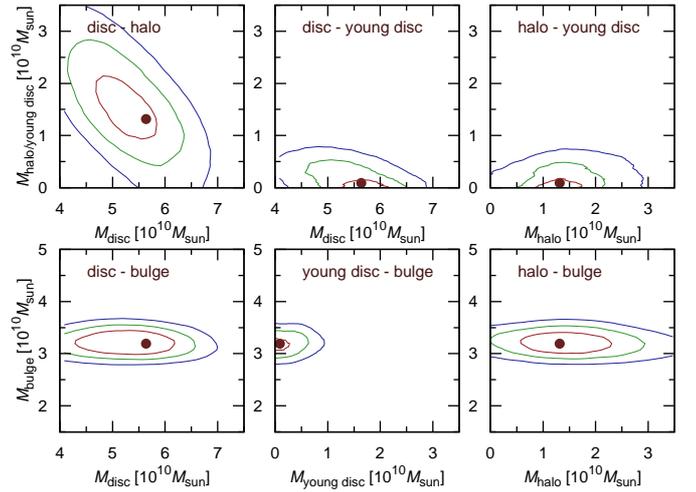}
\caption{Degeneracies between the masses of the different stellar components, shown as likelihood contours of various combinations of component masses. The final B07 model parameters are shown with dots.}
\label{fig:mass_like}
\end{figure}

It is natural to suspect that a four-component fit to the galaxy stellar mass distribution has to be degenerate to some extent. We tested the uniqueness of the model by varying the masses of the components and calculating the deviation of each resultant model from the observations using the Bayesian interface tool MultiNest \citep{Feroz:08, Feroz:09}. The results of the degeneracy analysis are presented in Fig.~\ref{fig:mass_like}, indicating the likelihood of different combinations of component masses. Quite expectedly, the most securely determined component is the bulge and the most unreliable mass estimates are for the young disc and halo, both being degenerate with the disc mass to some extent. The degeneracies would be higher, if all the component parameters were set free. A more conservative two-component (bulge + disc) model is described in Appendix~\ref{app:1}. 

%=================================================================
\section{Dynamics and dark matter distribution}\label{sect:dark_m}
%=================================================================

The structural model and masses of the stellar components derived in Sect.~\ref{sect:stellar_m} allowed us to calculate the gravitational potential of stellar matter in the galaxy. The gravitational potential of a galaxy is also traced by the rotation curve. To match the calculated rotation curve with the observed one, the contributions of gas and dark matter (DM) have to be added to the stellar mass model of the galaxy.

The contribution of gas to the potential of the galaxy is modest, thus a precise description of the gas distribution in the model is not essential. We have assumed that the distribution of the gas disc coincides with that of the young disc simply by raising the mass of the young disc by $6\cdot10^9M_{\odot}$, which is the approximate sum of the molecular \citep{Dame:93,Nieten:06} and the neutral \citep{carignan:06,chemin:09,corbelli:10} gas mass estimates. Note that the molecular gas content is rather low in M\,31, in fact even lower than the differences between the neutral gas mass estimates made by different authors.

We have considered various functional forms of DM density distribution while incorporating a DM halo component into the model galaxy. From observations of the dynamics of galaxies, distributions with a nearly constant inner density (and therefore, ``isothermal" or ``cored" haloes) have been derived, e.g. by \citet{burkert:95}:
\begin{equation}
\rho_\mathrm{Burkert}(r)=\frac{\rho_c}{\left(1+\frac{r}{r_\mathrm{c}}\right)
\left[1+(\frac{r}{r_\mathrm{c}})^2\right]} .
\end{equation}
On the other hand, N-body simulations suggest steeply rising DM density towards the centre (therefore ``cuspy" haloes, e.g. \citet{moore:99}:
\begin{equation}
\rho_\mathrm{Moore}(r)=\frac{\rho_\mathrm{c}}{(\frac{r}{r_\mathrm{c}})^{1.5}
\left[1+(\frac{r}{r_\mathrm{c}})^{1.5}\right]}
\end{equation}
and \citet{navarro:97} (hereafter NFW):
\begin{equation}
\rho_\mathrm{NFW}(r)=\frac{\rho_\mathrm{c}}{\left(\frac{r}{r_\mathrm{c}}\right)
\left[1+(\frac{r}{r_\mathrm{c}})\right]^{2}} .
\end{equation}
In these equations, $\rho_\mathrm{c}$ is a density scale parameter.

More recently, it has been found that the Einasto distribution Eq.~(\ref{eq:einasto}) matches with the simulated DM haloes over a wider range of radii \citep{merritt:06,navarro:10,chemin:11}, and is gaining popularity for various applications.

Each of the four DM distributions was used in combination with the stellar components as determined in Sect.~\ref{sect:stellar_m} to calculate the gravitational potential of the galactic model and the corresponding rotation curve. 

\begin{figure}
\includegraphics[width=88mm]{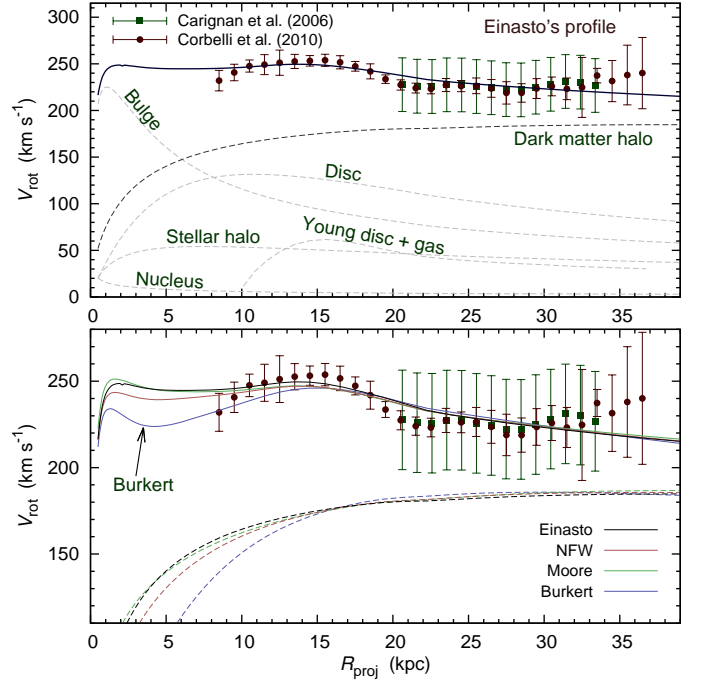}
\caption{\emph{Upper panel:} the observed rotation curve (data points with error bars) overplotted with the model (solid line). Contributions of each component are also shown (dashed lines). The model corresponds to the B07 stellar mass estimates and the Einasto distribution for the DM density. \emph{Lower panel:} the same stellar model with four different DM density distributions. For clarity, only the total rotation curves and the DM contributions are shown.}
\label{fig:rc}
\end{figure}

The model rotation curve was fitted to the observed rotation curve, composed of two \ion{H}{i} datasets from the literature: the Westerbork telescope observations \citep{corbelli:10} and data from the Effelsberg and Green Bank telescopes \citep{carignan:06}. We did not attempt to include observations from the inner parts of the galaxy, where the dynamics of gas clouds is too much affected by non-circular motions, leading to difficulties in interpreting the data.

In addition to the gas rotation curves, we used circular velocities calculated from the measurements of the motions of globular clusters, satellite galaxies, and stellar streams (Table~\ref{table:mass_estimates}), allowing us to trace the gravitational potential of M\,31 out to a projected distance of more than 500\,kpc from the centre.

For fitting the model rotation curve to the observed one, the DM halo parameters were left free, while the masses of the stellar components were kept fixed. During the first runs of the fitting, the Einasto shape parameter $N$ was allowed to vary freely, which lead to a wide variation of its value. To reduce degrees of freedom, we applied a fixed value at $N=6.0$ according to \citet{merritt:06} and \citet{navarro:10}.

%=========================================================
\begin{table}
    \caption{Enclosed mass estimates and the corresponding circular velocities at large galactocentric radii of M\,31.}
    \centering
    \begin{tabular}{lllll}
        \hline\hline
        $R$ & Mass & $V_\mathrm{c}$ & Objects & Reference \\
        $[\mathrm{kpc}]$ & $[$$10^{10}M_\odot]$ & $[$km\,s$^{-1}]$ & & \\
        \hline
        32  &  39$^{+2}_{-10}$  & 230$^{+5}_{-32}$  & 17 globular clusters & 1 \\
        37  &  49$^{+12}_{-14}$ & 240$^{+38}_{-38}$ & 21-cm data & 2 \\
        41  &  48$^{+34}_{-23}$ & 225$^{+69}_{-63}$ & 17 globular clusters & 1 \\
        55  &  55$^{+4}_{-3}$   & 208$^{+7}_{-6}$   & 504 globular clusters & 3 \\
        60  &  44$^{+26}_{-4}$  & 178$^{+48}_{-8}$  & 349 globular clusters & 4 \\
        100 &  79$^{+5}_{-5}$   & 185$^{+6}_{-6}$   & 12 satellites & 5 \\
        125 &  75$^{+25}_{-13}$ & 161$^{+25}_{-15}$ & stellar stream & 6 \\
        125 &  74$^{+12}_{-12}$ & 160$^{+13}_{-14}$ & stellar stream & 7 \\
        139 &  80$^{+41}_{-37}$ & 157$^{+36}_{-42}$ & 15 satellites & 8 \\
        268 & 137$^{+18}_{-18}$ & 149$^{+9}_{-10}$  & 7 satellites & 9 \\
        300 & 140$^{+40}_{-40}$ & 142$^{+19}_{-22}$  & satellites & 10 \\
        560 & 125$^{+180}_{-60}$&  98$^{+55}_{-27}$ & 11 satellites & 1 \\
        560 &  99$^{+146}_{-63}$&  87$^{+51}_{-34}$ & 16 satellites & 11 \\
        \hline
    \end{tabular}
    \tablebib{
    (1)~\citet{Evans:00}; (2)~\citet{corbelli:10}; (3)~\citet{Lee:08}; (4)~\citet{Galleti:06}; (5)~\citet{Cote:00}; (6)~\citet{Ibata:04}; (7)~\citet{Fardal:06}; (8)~\citet{Tollerud:12}; (9)~\citet{Courteau:99}; (10)~\citet{Watkins:10}; (11)~\citet{Evans:00a}.
    }
\label{table:mass_estimates}
\end{table}
%=========================================================

%=========================================================
\begin{table*}
    \caption{DM halo parameters for various distribution functions.}
    \centering
    \begin{tabular}{lllllll|ll}
        \hline\hline
        Profile & $\rho_c$ & $\rho_c$ & $r_c$ & $M_\mathrm{200}$ & $R_\mathrm{200}$ & $V_\mathrm{200}$ & $\rho_{-2}$ & $a_{-2}$ \\
                  & [$M_\odot$pc$^{-3}$] & [\mbox{GeV/$c^2$~cm$^{-3}$}] & [kpc] & $[10^{10}M_\odot]$ & [kpc] & [km\,s$^{-1}$] & [\mbox{GeV/$c^2$~cm$^{-3}$}] & [kpc] \\
        \hline
        Einasto\tablefootmark{a} 
                      & $8.12\pm 0.16\cdot 10^{-6}$ & $3.08\cdot 10^{-4}$ & $178\pm 18$ & 113 & 213 & 151 & $8.92\cdot 10^{-2}$ & 17.44 \\
        NFW           & $1.10\pm 0.18\cdot 10^{-2}$ & $4.18\cdot 10^{-1}$ & $16.5\pm 1.5$            & 104 & 207 & 147 \\
        Moore         & $1.46\pm 0.26\cdot 10^{-3}$ & $5.54\cdot 10^{-2}$ & $31.0\pm 3.0$            & 106 & 209 & 148 \\
        Burkert       & $3.68\pm 0.40\cdot 10^{-2}$ & $1.40\cdot 10^{0}$ & $9.06\pm 0.53$            & 79  & 189 & 134 \\
        \hline
        Einasto\tablefootmark{a}\tablefootmark{b}
                      & $1.40\pm 0.27\cdot 10^{-6}$ & $5.32\cdot 10^{-5}$ & $387\pm 44$ & 127 & 221 & 157 & $1.54\cdot 10^{-2}$ & 37.95 \\
        \hline
    \end{tabular}
    \tablefoot{
    \tablefoottext{a}{Parameter $N$ has been taken 6.0, yielding $d_N=17.668$. Spherical symmetry is assumed by taking $q=1$ in Eq.~(\ref{eq:einasto}), in which case $r_c = a_c$.}
    \tablefoottext{b}{Dark matter parameters for the maximum-stellar model.}
    }
\label{table:dark_comp}
\end{table*}
%=========================================================

As shown in Sect.~\ref{sect:mod}, the stellar masses yielded by different stellar population synthesis models vary by a factor of two. We first considered two cases: the lowest-mass model with the B07 mass estimates and the highest-mass model with the other mass estimates. In the latter case, however, the stellar mass becomes too high, raising the model rotation curve above the observed values at distances 10--20~kpc from the centre even without the inclusion of a dark matter component (a DM halo would still be needed though to gain a match with the outer rotation curve observations). Thus we had to abandon the idea of determining the upper limits of the masses of the stellar components by using stellar population synthesis models.

In cases when stellar masses cannot be estimated independently, the maximum-disc approach is often followed, i.e. first a pure disc is fitted to the observed rotation curve with as high mass as possible and the other components are added thereafter in the required proportions. In the case of M\,31, the disc mass-to-light ratio is very close to the bulge one because of their similar ages and metallicities. Therefore, instead of a maximum-disc approach, we applied a maximum-stellar model, conserving the relative values of the mass-to-light ratios of the stellar components as determined with stellar population synthesis models, but multiplying them with a common constant. Without the inclusion of a DM halo, the rotation curve allowed the multiplication of the B07 stellar masses bye 1.5 at maximum. Henceforth, we are using the corresponding model (together with a minimally required DM halo) as an upper limit of the stellar masses and refer to it as the maximum stellar model.

Parameters of the best-fitting DM models, corresponding to the different distribution functions and the B07 stellar mass estimates are presented in Table~\ref{table:dark_comp}. For the Einasto DM distribution, parameters corresponding to the maximum-stellar model are also given.

\begin{figure}
\includegraphics[width=88mm]{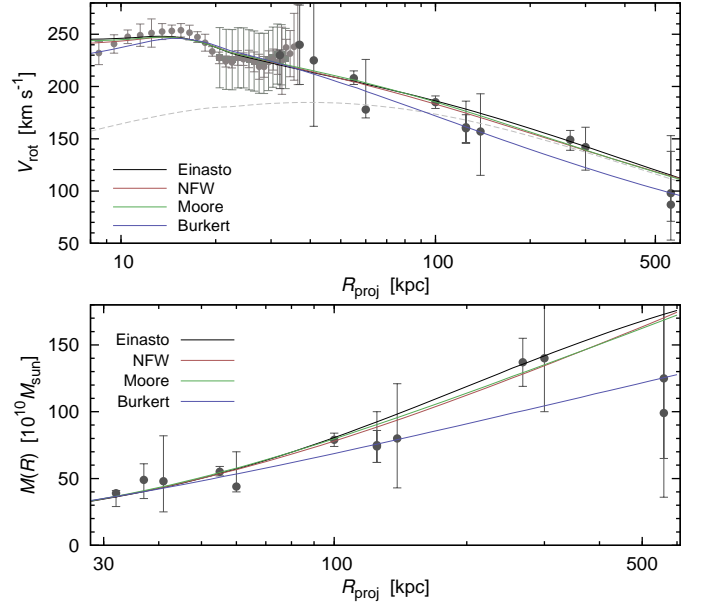}
\caption{Outer rotation curve observations and models (\emph{upper panel}), calculated for the B07 stellar masses, and the corresponding cumulative mass (\emph{lower panel}). With the exception of the Burkert distribution, all DM models fit well to the observations.}
\label{fig:tot_mass}
\end{figure}

The upper panel of Fig.~\ref{fig:rc} presents the observed rotation curve, over-plotted with the curve derived from the B07 stellar masses and the Einasto DM halo. Contributions of each stellar component and the DM halo are also shown. In the lower panel, model rotation curves for all the four DM models are presented. It is seen that within the observed range of the rotation curve, differences between different DM profiles are negligible. From 7\,kpc inwards along the major axis, outside the range of observations, the model with the Burkert DM profile (and to a lesser extent, also the model with the NFW DM profile) starts to deviate from the other models. In Fig.~\ref{fig:tot_mass}, the outer rotation curve (upper panel) and the corresponding enclosed mass (lower panel) are shown together with the model curves. Again, all the DM distribution models match the observations within the errorbars, except for the Burkert distribution, which produces a slightly lighter DM halo, missing a few outer datapoints.

\begin{figure}
\includegraphics[width=88mm]{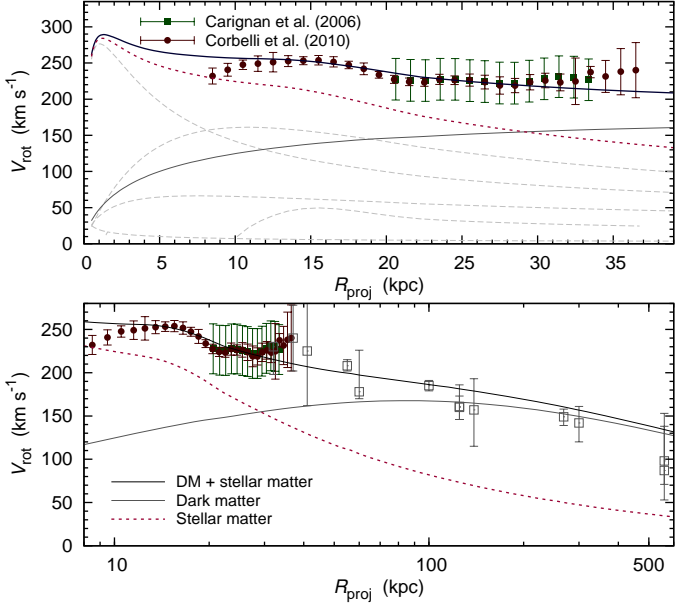}
\caption{The observed rotation curve together with the maximum-stellar model, in which the stellar masses are 1.5 times higher than in the B07 model.}
\label{fig:rc_stellar}
\end{figure}

The model rotation curves for the maximum-stellar model are shown in Fig.~\ref{fig:rc_stellar}. Now the fit is somewhat worse than for the B07 model, especially at the innermost observational datapoints, securing that the maximum-stellar model indeed provides the very upper limits for the stellar masses.

\begin{figure}
\includegraphics[width=88mm]{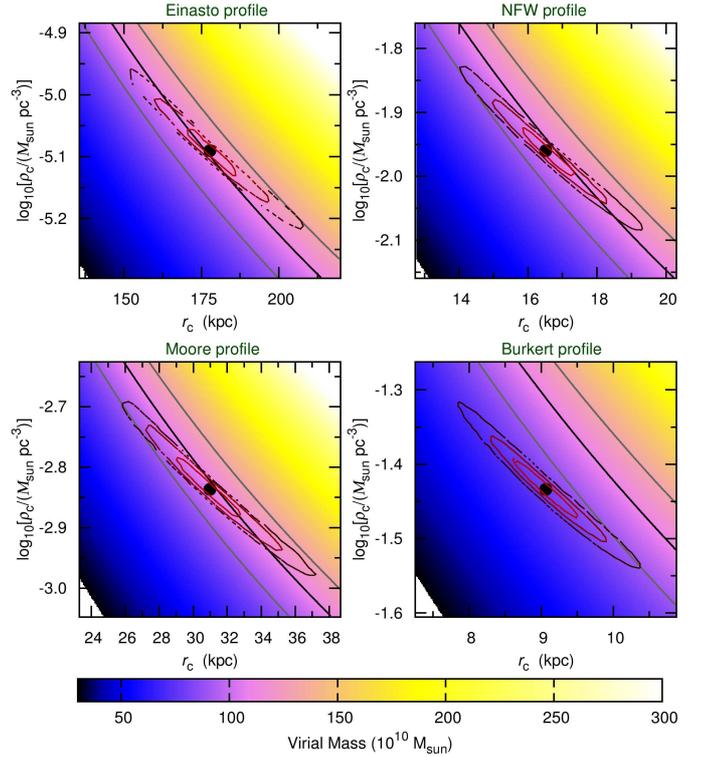}
\caption{Parameter likelihoods for different DM density distributions in the case of B07 stellar masses. The virial mass corresponding to each parameter combination is shown in colour coding; 90$\cdot10^{10}M_{\odot}$, 110$\cdot10^{10}M_{\odot}$, and 130 $\cdot10^{10}M_{\odot}$ levels are indicated with solid contours. For the Einasto, NFW, and Moore DM profile, the virial mass is almost the same.}
\label{fig:dm_params}
\end{figure}

\begin{figure}
\includegraphics[width=88mm]{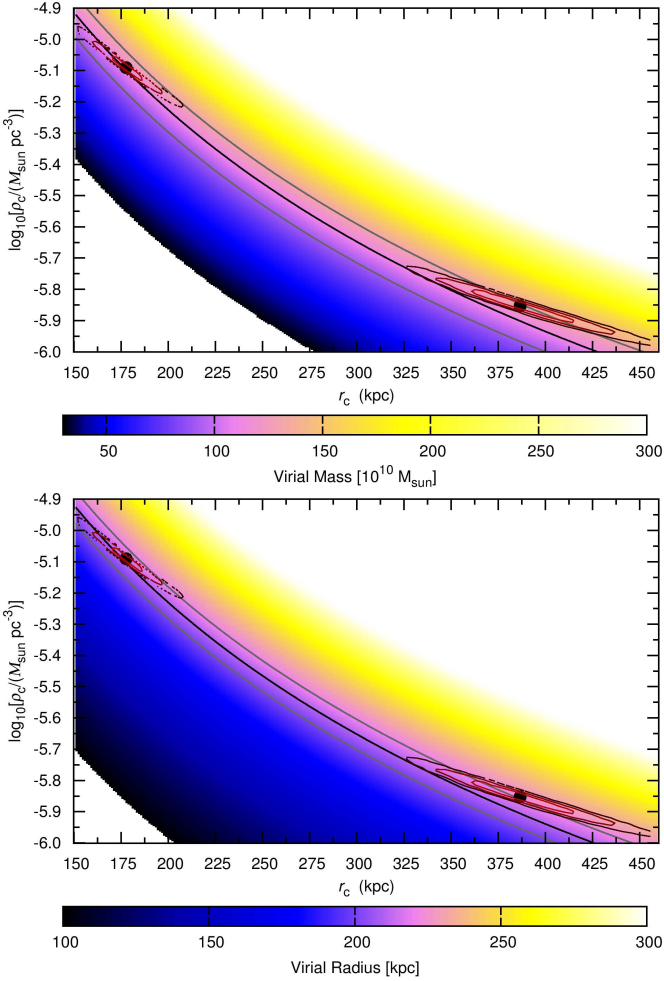}
\caption{Comparison of DM halo parameters for B07 and maximum stellar mass models. In the upper panel, the virial mass corresponding to each parameter combination is shown in colour coding; 90$\cdot10^{10}M_{\odot}$, 110$\cdot10^{10}M_{\odot}$, and 130 $\cdot10^{10}M_{\odot}$ levels are indicated with solid contours. In the lower panel, virial radii are shown in colour coding, with solid contours tracing 200, 210, and 220~kpc.}
\label{fig:app_dm_params}
\end{figure}

As shown above, it is not possible to prefer any of the given DM distribution models on the basis of the data on M\,31. Furthermore, in each case, the derived characteristic radii and densities of the DM haloes are very degenerate, as indicated in Fig.~\ref{fig:dm_params}: a significant increase of the characteristic radius can easily be compensated by lowering the characteristic density and vice versa. In this plot, the virial mass $M_\mathrm{200}$, defined as the mass within a sphere of mean density 200 times the cosmological critical density\footnote{Here the critical density is calculated for the Hubble constant value $H_0=71\,$km\,s$^{-1}$Mpc$^{-1}$.}, \citep[e.g.][]{navarro:10}, is also shown in colour coding for each DM model. Interestingly, despite the uncertainty of the DM density distribution, the virial mass is well constrained, regardless of the chosen DM distribution model. The same statement holds in Fig.~\ref{fig:app_dm_params}, where the Einasto DM halo parameter likelihoods for the two stellar mass models are compared. The virial mass is actually quite firmly established by the outer ``test particles" of the dynamics and is almost independent of the stellar model of the galaxy. For the ``cuspy" DM profiles (Einasto, NFW, Moore), the derived virial mass is $(1.04$--$1.13)\cdot10^{12}M_{\odot}$ and the corresponding virial radius $207$--$213$\,kpc. For the ``cored" Burkert profile, the values are $0.8\cdot10^{12}M_{\odot}$ and $189$\,kpc, respectively. In the case of the Einasto DM distribution, a sphere extending to 10 pc from the centre contains 16--61\,$M_{\odot}\mathrm{\,pc^{-3}}$ (0.6--2.3\,$\mathrm{TeV}/c^2\,\mathrm{cm}^{-3}$) of DM on an average.

%=========================================================
\section{Discussion and comparison to previous models}
%=========================================================

%============================================================================
\begin{table*}
    \caption{Comparison of bulge, disc, and DM halo mass estimates. All masses are in $10^{10}M_\odot$.}
    \label{table:mass_comp}
    \centering
    \begin{tabular}{lccc}
        \hline\hline
        Model &  $M_{\mathrm{bulge}}$ &  $M_{\mathrm{disc}}$   &  $M_{200}$  \\
        \hline
        \citet{geehan:06}, best-fit (maximum-disc) model    & 3.3       & 8.4 (13.7) &  68 (94)                 \\
        \citet{seigar:08}, model without (with) adiabatic contraction   & 3.5 (3.5) & 5.8 (7.3)  & 73 (89)  \\
	\citet{chemin:09}, ``hybrid" model    & 2.32      & 7.1        &  100                     \\
	\citet{corbelli:10}, NFW model with $(M/L)_\mathrm{bulge} = (M/L)_\mathrm{disc}$  & 3.8 & 8.8 &  85\tablefootmark{c} \\
	This work, B07 model                            & 4.4\tablefootmark{a}   &  5.7\tablefootmark{b}   &  113              \\
	This work, maximum-stellar model		      & 6.6\tablefootmark{a} & 8.6\tablefootmark{b} &  127\\
        \hline
    \end{tabular}
    \tablefoot{
    \tablefoottext{a}{Sum of the bulge and stellar halo masses.}
    \tablefoottext{b}{Sum of the disc and young disc masses.}
    \tablefoottext{c}{Recalculated from $M_{98}$ in the original paper.}
    }
\end{table*}
%=============================================================================

\begin{figure}
\includegraphics[width=88mm]{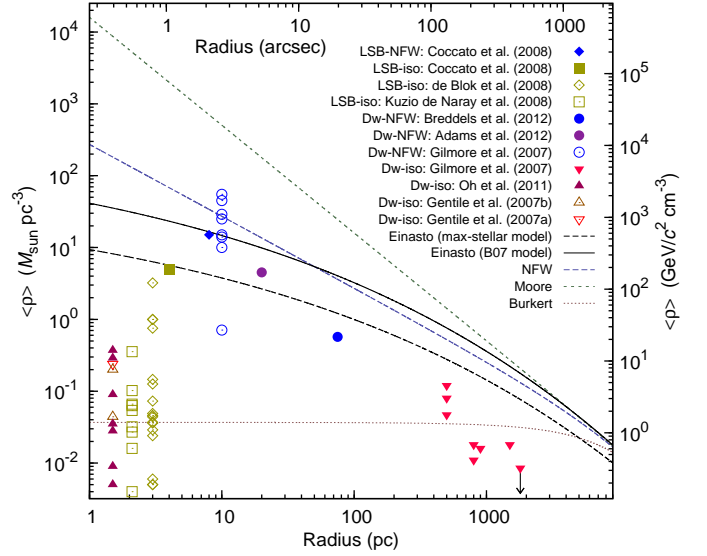}
\caption{Average DM density inside a given radius, corresponding to different DM distributions in the case of the B07 stellar masses. For the Einasto DM distribution, also the maximum-stellar-mass case is plotted. For comparison, central densities of some nearby dwarf galaxies and low-surface-brightness galaxies are shown. The triangular/quadrangular datapoints are calculated assuming the Burkert DM, the circular datapoints correspond to the NFW DM.}
\label{fig:dm_cumul}
\end{figure}

Let us compare our mass models of M\,31 to some other recently constructed models. In Table~\ref{table:mass_comp}, mass estimates suggested by our models for the bulge, disc, and DM halo are compared to the estimates by \citet{geehan:06}, \citet{seigar:08}, \citet{chemin:09}, and \citet{corbelli:10}. For a better understanding of the compatibility of these models, we will briefly summarise the principal properties and differences of these models below.

The model derived by \citet{geehan:06} consists of a central supermassive black hole, a bulge, a disc, and a DM halo. Its stellar components are determined using various luminosity measurements in $V$, $R$, and $r$ filters out to the (projected) distance of 25\,kpc along the major axis, and the kinematics is based on a composite rotation curve. Mass-to-light ratios of the stellar components are treated as free parameters. The total mass is constrained by data on the motions of outer planetary nebulae, globular clusters and satellite galaxies. In Table~\ref{table:mass_comp}, \citet{geehan:06} values for the best-fit model and the maximum-disc model (in brackets) are presented.

\citet{seigar:08} constructed a similar black hole\,+\,bulge\,+\,disc\,+\,DM halo model. They used the Spitzer 3.6-micron imaging data and \mbox{$B\!-\!R$} colour profile to determine the mass-to-light ratios; dynamical mass estimators were the same or similar as in \citet{geehan:06}. In addition to the usual DM profiles, \citet{seigar:08} considered the case of a dark halo that has undergone an adiabatic contraction due to the gravitational attraction of the baryonic material. In Table~\ref{table:mass_comp}, the model with adiabatic contraction is presented in brackets; it should not be compared directly to the other models.

\citet{chemin:09} constructed a black hole\,+\,bulge\,+\,disc\,+\,gas model, using their newer \ion{H}{i} data for constraining the kinematics. For a more accurate description of the disc potential, the disc density distribution was settled as the residual of the surface brightness distribution after the subtraction of the bulge contribution. The small contribution of gas mass was considered on the basis of $\mathrm{H}_2$ and \ion{H}{i} surveys. In Table~\ref{table:mass_comp}, the ``hybrid" model (with the bulge mass is determined from stellar velocity dispersions and the disc mass from simple stellar population models) values of \citet{chemin:09} are presented.

\citet{corbelli:10} used a bulge\,+\,disc\,+\,gas model together with an optional Burkert/NFW DM halo to fit their \ion{H}{i} kinematics data and outer dynamics estimators. The best-fit model with equal bulge and disc mass-to-light ratios and the NFW dark halo is shown in Table~\ref{table:mass_comp}. We have rescaled the virial mass $M_{98}$ given by \citet{corbelli:10} to $M_{200}$.

In contrast to these four works, our stellar model is based on fully 2-dimensional dust-corrected imaging through 6 filters and some more recent dynamical mass estimators. Instead of the central black hole, our model includes the nucleus of the galaxy, which contributes significantly more to the total mass and the model rotation curve of the galaxy. This contribution is nevertheless tiny and has a negligible effect on other model parameters, as well as our usage of an oblate bulge \citep[with an axial ratio 0.8;][]{tempel:11} instead of a spherical one. In Table~\ref{table:mass_comp}, our B07 model and the maximum-stellar model results are given. The actual values probably lie between the estimates of these two models.

Table~\ref{table:mass_comp} shows that the bulge mass suggested by our models is somewhat higher than in previous models, probably resulting from different bulge parameters but also because we have used a larger set of observational data to constrain the stellar populations. Nevertheless, the bulge dominates the total gravitational potential only up to the radius 6--8\,kpc and bulge properties have little effect on the DM halo parameters.
 
As can be seen from Table~\ref{table:mass_comp}, the DM halo virial mass estimates have previously remained mostly below 1$\cdot10^{12}M_\odot$, whereas our models suggest slightly higher values, (1.0--1.3)\,$\cdot10^{12}M_\odot$. Once again, the most likely source of differences is our usage of a larger collection of observational data on the outer dynamics. As shown in Figs.~\ref{fig:dm_params} and~\ref{fig:app_dm_params}, the virial mass is almost independent of the DM density profile and the stellar mass model, being uniquely determined by the outer dynamics of the galaxy.

In Fig.~\ref{fig:dm_cumul}, the distribution of the average DM halo density within a given radius is shown for each DM density distribution model. The added datapoints provide an illustrative comparison with DM haloes of other galaxies, for which the average central density has been recently measured more or less reliably: nearby dwarf galaxies \citep{Gilmore:07, Gentile:07a, Gentile:07b, Oh:11, Adams:12, Breddels:12} and low-surface-brightness galaxies \citep{Coccato:08, deBlok:08, KuziodeNaray:08}. Some of these values are calculated for the modified-isothermal DM distribution, others for the NFW distribution. In several cases, both versions are presented. These datapoints should be compared to the Burkert and NFW profiles of M\,31, respectively. It is seen that the central density of DM haloes varies by a couple of orders of magnitude and despite its higher total mass, the DM halo of M\,31 cannot be distinguished from an average dwarf or low-surface-brightness galaxy in this aspect. Interestingly, the estimate of the central density (0.012--0.028)$\cdot M_{\odot}\mathrm{pc^{-3}}$ of DM haloes of massive disc galaxies near redshift $z\simeq0.9$ \citep{tamm:05} also falls within this range, hinting that the DM halo concentration process seems to be restricted rather uniformly over a very wide variety of halo masses, environments, and cosmological epochs. 

To conclude our work, it is interesting and also disappointing to note that the usage of additional observational data does not reduce significantly the uncertainties and scatter of the parameters of M\,31 mass distribution models. Our vague understanding of the evolution of the physical properties of stellar populations restrains the gain from all the gathered observational information on the chemical content and formation history of a galaxy. Despite the improved imaging and kinematics data, we are still unable to confirm or rule out the maximum-disc or maximum-baryonic approach in splitting the contributions of luminous and dark matter to the overall mass distribution. It can only be concluded that the bulge mass of M\,31 probably lies within the range (4.4--6.6)\,$\cdot10^{10}M_\odot$ and the disc mass within the range (5.7--8.6)\,$\cdot10^{10}M_\odot$. Nevertheless, M\,31 provides an exceptional opportunity to estimate the virial mass and the outer distribution of a DM halo thanks to the possibility of tracing the gravitational potential with various test bodies at large distances from the galactic centre.

%=========================================================
\section*{Acknowledgments}
%=========================================================

We are grateful to the anonymous referee for carefully reading the manuscript and for making useful suggestions for improving it. We acknowledge the financial support from the Estonian Science Foundation (incl. the grant MJD272)  and the Ministry of Education and Research.

This research has made use of the NASA/ IPAC Infrared Science Archive, which is operated by the Jet Propulsion Laboratory, California Institute of Technology, under contract with the National Aeronautics and Space Administration. All the figures were made with the gnuplot plotting utility.

%\bibliographystyle{aa}%
%\bibliography{5t}{}

%=========================================================

%=========================================================

\begin{appendix}

%=========================================================
\section{Simple bulge\,+\,disc model of M\,31} \label{app:1}
%=========================================================

\begin{figure}
\includegraphics[width=88mm]{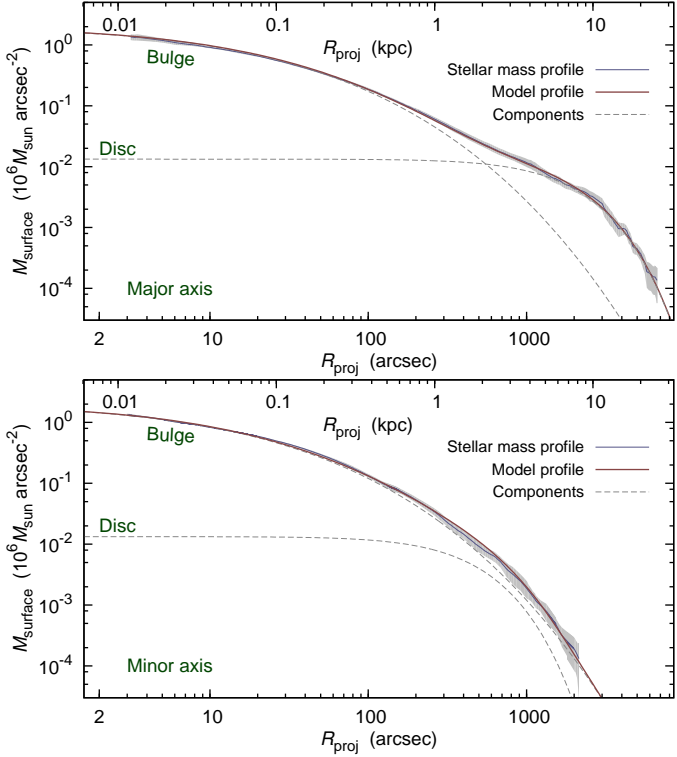}
\caption{Mass profile for two stellar components. For major (\emph{upper panel}) and minor (\emph{lower panel}) axis.}
\label{fig:app_profile}
\end{figure}

\begin{figure}
\includegraphics[width=88mm]{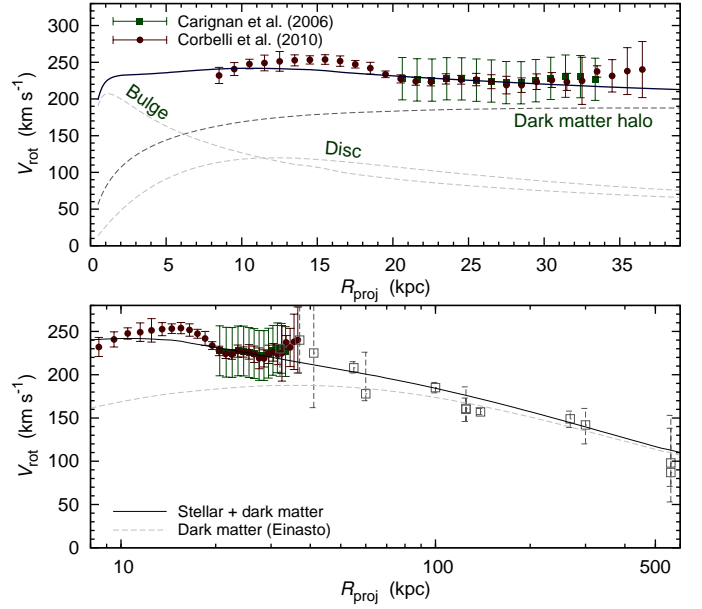}
\caption{The observed inner (\emph{upper panel}) and outer (\emph{lower panel}) rotation curve compared to the model with two stellar components.}
\label{fig:app_rc}
\end{figure}

For several applications (e.g. for comparing to models of more distant galaxies), a model of M\,31 with two stellar components is accurate enough. In Fig.~\ref{fig:app_profile} we present such a model, conservatively taking $N=4$ for the bulge and $N=1$ for the disc. The other parameters are determined by fitting the model components to the B07 stellar mass density distribution derived in Sect.~\ref{sect:stellar_m}; they are presented in Table~\ref{table:two_comp}. The corresponding rotation curve, including the contribution of the Einasto DM halo as derived above, is compared to the observations in Fig.~\ref{fig:app_rc}. 

%=========================================================
\begin{table}
    \caption{Parameters for the two-component model.}
    \label{table:two_comp}
    \centering
    \begin{tabular}{lcccccc}
        \hline\hline
        Comp. & $a_c$ & $q$ & $N$ & $d_N$ & $\rho_c$ & $M_\mathrm{comp}$  \\
                  & [kpc] &     &     &       & ${[M_\odot\,\mathrm{pc}^{-3}]}$ & $[10^{10}M_\odot]$ \\
        \hline
        Bulge & 2.025  & 0.73 & 4.0 & 11.67 & $2.20\cdot 10^{-1}$ & 4.9   \\
        Disc  & 11.35 & 0.10 & 1.0 &  2.67 & $1.72\cdot 10^{-2}$ & 4.8   \\
        \hline
    \end{tabular}
\end{table}

%================================================================================
\section{Relations between different forms of Einasto distribution} \label{app:2}
%================================================================================

In our previous papers \citep{tempel:06,tempel:10,tempel:11}, we have used the Einasto distribution function in the form as originally defined and used by Jaan Einasto \citeyearpar{Einasto:69} and thereafter, in which the density distribution is given in respect to the harmonic mean radius $a_0$. Galaxy component is approximated by an ellipsoid of revolution with a constant axial ratio $q$; its spatial density distribution follows the law
\begin{equation} 
    \rho(a)=\rho_0\exp \left[ -\left( {a \over ka_0}\right)^{1/N}\right] , \label{eqapp:eq1}
\end{equation}
where $\rho_0=hM/(4\pi q a_0^3)$ is the central density and $M$ is the component mass; $a=\sqrt{r^2+z^2/q^2}$, where $r$ and $z$ are two cylindrical coordinates. The coefficients $h$ and $k$ are normalising parameters, dependent on the structure parameter $N$.

A detailed derivation of the constant $h$ and $k$ depending on $N$ are given in Appendix~B of \citet{tenjes:94}. In short, to find $h$ and $k$, the following equality must be satisfied
\begin{equation}
    h^{-1}\equiv\int\limits_0^\infty x \exp\left[ -\left(\frac{x}{k}\right)^{1/N} \right] \mathrm{d}x =
    \int\limits_0^\infty x^2\exp\left[ -\left(\frac{x}{k}\right)^{1/N} \right] \mathrm{d}x .
\end{equation}
These integrals can be solved analytically, giving
\begin{equation}
k=\frac{\Gamma(2N)}{\Gamma(3N)}, \qquad
h=\frac{\Gamma^{2}(3N)}{N\Gamma^{3}(2N)},
\label{eq:hk}
\end{equation}
where $\Gamma$ is the (complete) gamma function.

The advantages of this form are the usage of the harmonic mean radius as a good characteristic of the real extent of the component, rather independent of the structural parameter $N$, and the directly calculable integral mass (or total luminosity).

More recently, instead of the usual Navarro-Frenk-White (NFW) formula \citep{navarro:97}, the Einasto law has been used for DM haloes \citep{navarro:04,merritt:06}. In these works, the equation is used in the form
\begin{equation}\label{eqapp:eq2}
\rho(a)= \rho_c\exp\left\{-d_{N}
\left[\left(\frac{a}{a_{c}}\right)^{1/N}-1\right]\right\},
\end{equation}
where $d_N$ is a function of $N$, such that $a_c$ defines a volume containing half of the total mass and $\rho_c$ becomes the density at radius $a_c$; $a$ is the same as in Eq.~(\ref{eqapp:eq1}). This Eq.~(\ref{eqapp:eq2}) is in fact the same as used by J. Einasto in his first paper \citep{Einasto:65} to describe the spatial density distribution of galactic components.

An integral of Eq.~(\ref{eqapp:eq2}) over some volume gives the enclosed mass, which is finite and in the spherical case ($q=1$), equal to
\begin{equation}\label{eqapp:mass}
M(a)=4\pi\int\limits_0^{a}\rho(x)x^2\mathrm{d}x=
4\pi N a_c^3 \rho_c \frac{e^{d_N}}{d_N^{3N}} \gamma\left[3N,d_N \left(\frac{a}{a_c}\right)^{\frac{1}{N}}\right],
\end{equation}
where $\gamma$ is the lower incomplete gamma function.

Replacing $\gamma[3N,d_N (a/a_c)^{1/N}]$ with $\Gamma(3N)$ in Eq.~\ref{eqapp:mass} gives the total mass of the component.

The value of $d_N$ can be found exactly, by solving %$M(\infty)=2M(a_c)$.
$\Gamma(3N)=2\gamma(3N,d_N)$. \citet{merritt:06} showed that the value of $d_N$ can be approximated by the expression
\begin{equation}\label{eqapp:dn}
d_N  \approx 3N-1/3+0.0079/N \quad \mathrm{for}\quad N\ga0.5.
\end{equation}
A slightly more exact approximation for $d_N$ is derived in \citet{Retana-Montenegro:12}, applicable also for $N\ga0.5$.

\begin{figure}
\includegraphics[width=88mm]{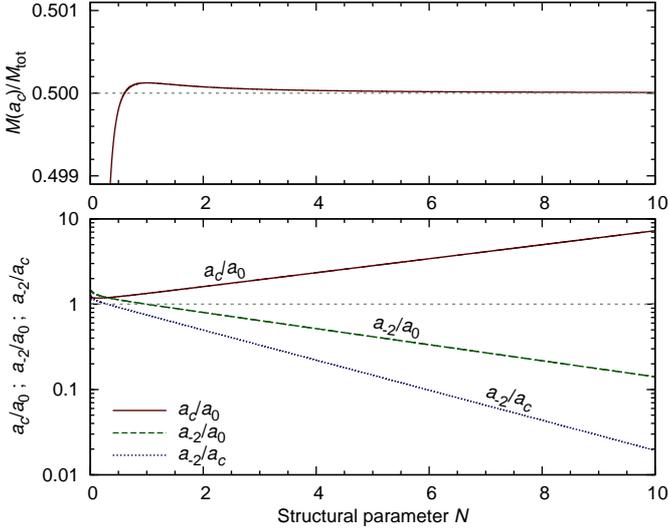}
\caption{\emph{Upper panel} shows the mass fraction for a radius $a_c$ as a function of $N$, where $d_N$ is found using the Eq.~(\ref{eqapp:dn}). Note the high amplification level of the y-axis scale. \emph{Lower panel} shows the relation between harmonic mean radius $a_0$, half-mass radius $a_c$, and radius $a_{-2}$; $d_N$ is calculated using the Eq.~(\ref{eqapp:mass}).}
\label{fig:app_teisendus}
\end{figure}

In the upper panel of Fig.~\ref{fig:app_teisendus} we show the mass fraction within radius $a_c$ as a function of $N$, where $d_N$ is found from Eq.~(\ref{eqapp:dn}). It is seen that for $N\ga0.5$, Eq.~(\ref{eqapp:dn}) gives a good approximation. For lower $N$ values, the exact solution should be preferred for $d_N$. Throughout this work, we have used and presented the exact $d_N$ values.

In addition to Eqs.~(\ref{eqapp:eq1}) and (\ref{eqapp:eq2}), the Einasto distribution for DM haloes has been used in the form
\begin{equation}
    \label{eqapp:eq3}
\rho(a)= \rho_{-2}\exp\left\{ \frac{-2}{\alpha}
\left[\left(\frac{a}{a_{-2}}\right)^{\alpha}-1\right]\right\},
\end{equation}
where $\alpha=1/N$ and $a_{-2}$ marks the radius where the logarithmic slope of the profile\footnote{In fact, in the original paper \citet{Einasto:65} introduced just this kind of formula. Instead of $\rho_{-2}$ he used a logarithmic slope in the solar neighbourhood.} equals the isothermal value \mbox{$-\mathrm{d}\ln{\rho}/\mathrm{d}\ln a=2$}. This form of the Einasto law is gaining popularity for describing DM haloes in $N$-body simulations \citep[e.g.][]{navarro:10}, as well as applications in astroparticle physics \citep[e.g.][]{Hutsi:10,Tempel:12a}. 

All forms of the Einasto law (Eqs.~\ref{eqapp:eq1}, \ref{eqapp:eq2}, and \ref{eqapp:eq3}) are identical, and strict transductions exist between the parameters:
\begin{eqnarray}
    a_c&=&a_0 k \left(d_N\right)^N ,\\
    \rho_c&=&\rho_0 \exp\left({-d_N}\right) \label{eqapp:lc} ,\\
    a_{-2}&=&a_0 k (2N)^N = a_c\left(2N/d_N\right)^N ,\\
    \rho_{-2}&=&\rho_0\exp\left(-2N\right) = \rho_c\exp\left(d_N-2N\right) .
\end{eqnarray}

The lower panel in Fig.~\ref{fig:app_teisendus} shows the relation between the harmonic mean radius $a_0$, half-mass radius $a_c$, and radius $a_{-2}$. It is seen that with increasing $N$ the profile becomes more extended and also the half-mass radius increases, while $a_{-2}$ decreases compared to the harmonic mean radius $a_0$.

Observations have demonstrated that the stellar disc can have a toroidal form in some galaxies, i.e. it does not continue to the centre (one example is currently studied galaxy, M\,31). One simple way to represent a stellar population with the central density minimum in the presented density distribution framework was introduced by J.~Einasto \citep{Einasto:69,Einasto:80}. The spatial density of a disc with a central hole can be expressed as a sum of two spheroidal mass distributions
\begin{equation}
    \label{eq:lumhole}
    \rho_{\mathrm{disc}}(a) = \rho_{+}(a) + \rho_{-}(a) ,
\end{equation}
both of which can be approximated with the exponential law (\ref{eqapp:eq1}). Characterising the extent of the minimum with a parameter $\kappa=a_{0-}/a_{0+}$ and adopting a model disc with a zero density at $r=0$ and a non-negative density at $\rho_{\mathrm{disc}}(a) > 0$, the following relations must hold: $M_{-}=-\kappa^2M_{+}$, $q_{-}=q_{+}/\kappa$, where $\kappa < 1$ is a parameter that determines the relative size of the hole in the centre of the disc. The structural parameters $N_{-}$ and $N_{+}$ should be identical.

\end{appendix}

\end{document}